# Modelling of Heat Transfer in Single Crystal Growth

Alexander I. Zhmakin

Ioffe Physical Technical Institute, Russian Academy of Sciences, St. Petersburg, Russia
Softimpact Ltd., P.O. 83, 194156 St. Petersburg, Russia
Email: ai@softimpact.ru

**ABSTRACT** An attempt is made to review the heat transfer and the related problems encountered in the simulation of single crystal growth. The peculiarities of conductive, convective and radiative heat transfer in the different melt, solution, and vapour growth methods are discussed. The importance of the adequate description of the optical crystal properties (semitransparency, specular reflecting surfaces) and their effect on the heat transfer is stresses. Treatment of the unknown phase boundary fluid/crystal as well as problems related to the assessment of the quality of the grown crystals (composition, thermal stresses, point defects, disclocations etc.) and their coupling to the heat transfer/fluid flow problems is considered. Differences between the crystal growth simulation codes intended for the research and for the industrial applications are indicated. The problems of the code verification and validation are discussed; a brief review of the experimental techniques for the study of heat transfer and flow structure in crystal growth is presented. The state of the art of the optimization of the growth facilities and technological processes is discussed. An example of the computations of the heat transfer and crystal growth of bulk SiC crystals by the sublimation method is presented.

## INTRODUCTION

Computational Heat Transfer (CHT), being a part of modern technology, heavily depends on the availability of high quality single crystals. Making of a processor chip starts with cutting a substrate from a bulk Si crystal, then proceeds to photolithography using large $CaF_2$ crystals as lenses. Oxide crystals such as $Bi_4Ge_3O_{12}$ are used in the medical tomographic equipment. Soon GaInN based light emitting diodes will provide lighting. Solid-state devices are present in mobile phones used to inform the boss on advances in heat transfer simulation or just to reserve a table at the restaurant to celebrate the progress. Those who dislike straight whisky need ice crystals. And last but not he least, the better part of the computer scientists wear jewelry while the rest need gems grown either in the nature or in the laboratory to give a wife or a girlfriend. Evidently, CHT should in turn help to grow better crystals [1].

Industrial crystal production has been started by Verneuil in 1902 using flame-fusion to grow ruby and sapphire with melting points above 2300 K; now the world crystal production is estimated at more than 20 000 tons per year [2]. The growth of crystals involves a number of physical processes, the crucial one usually being heat transfer; solidification of the pure substance is driven by the temperature evolution only. There are a few exceptions such as epitaxial growth in "hot wall" reactors [3], protein crystallization frequently driven by the concentration gradients under isothermal conditions or growth of ultra-thin InGaAsP layers by liquid phase epitaxy onto the fast moving substrate [4]. Majority of inorganic crystals (semiconductors, laser and nonlinear crystals, optical crystals, jewelry, scintillation crystals) are grown at high temperatures and thus radiative heat transfer is often a dominant mechanism. Still, heat transfer via conduction and convection are equally important since they determine local temperature fields at the growth surface and thus to a great extent both the growth rate and the morphological stability. A relative role of different mechanisms depend on a number of factors including crystal scale, fluid medium used (vapour, solution, melt) and the growth method. To get an impression of the range of scales in crystal growth, one can compare the dimensions of the facilities for two close relatives: an ammonothermal method for bulk GaN growth from supercritical ammonia (diameter of the reactor tube is 3mm, its length is 0.18 m [5]) and a hydrothermal quartz growth method (an autoclave has a diameter of 650 mm and a height of 14 m and provides yield up to 3000 kg in a single run [6]).

The aims of numerical simulation are to explain and to predict. Numerical simulation is not a substitute for experiment, but a complement to it. Numerical models can provide detailed information on the flow, temperature and concentration fields, strain in the crystals etc. which can be measured experimentally only partly or not at all. On the other hand, numerical models depend on experimental data (materials properties, boundary conditions and so on). Moreover, numerical predictions are unreliable unless models are validated using experimental data in the widest possible range of macroscopic parameters.

"Modelling" and "Simulation" are frequently used as synonims. In CHT and Computational Fluid Dynamics (CFD) community, however, the former usually refers to the development or modification of a model while the latter is reserved for the application of the model [7]. A model should relate the process specification (equipment geometry, materials properties and transport coefficients, techological parameters such as pulling rate, the heating power and heater position, crystal and crucible rotation rates, input concentrations and mass flow rates, external electromagnetic fields, orientation of the growth facility etc.) to its outputs: crystals yield, thickness and composition uniformity, crystal quality, process duration and production costs. The list of the quality characteristics, being material depending, could include point defects (vacancies, interstitials); dislocations; twins and low angle boundaries for binary $A_3B_5$ compound semiconductors such as GaAs, InP; micropipes and polytype inclusions for silicon carbide; voids and oxygen precipitates for silicon; microbubbles for optical crystals; micro- and macrocracks for ternary semiconductor crystals.

The straightforward use of the model is referred to as a *direct* problem. From a practical point of view the reversed formulation is more useful: how one should change the equipment design or the process parameters to improve the crystal quality or to reduce production costs, for example. The simplest way is a "try-and-error" approach: to use one's intuition to introduce changes in the process specification, perform simulation and evaluate results. Sometimes, especially at the early development stages, an even simpler "blind search" [8] approach is exploited which essentially is a screening of a range of parameters. A more systematic way is to state an *inverse* problem by indicating
   1) which geometry characteristics or operating conditions (control parameters) could be varied;
   2) what criteria should be used to measure the success of the optimization.
The difficulty of the solution of inverse problems is their ill-posedness [9].

In the present paper an attempt is made to review the problems of numerical simulation of heat transfer in the industrial crystal growth and approaches to their solution. The growth of single crystals only will be considered; an important and complex problem of simulation of dendritic solidification is beyond the scope of this paper (an introduction to this problem and review of the corresponding numerical methods can be found, for example, in refs. [10,11]).

## MATHEMATICAL MODELS AND NUMERICAL METHODS

Mathematical models, as well as numerical methods, used for simulation of crystal growth are essentially the same as in other Computational Continuum Mechanics (CCM) applications (heat transfer, fluid dynamics, electromagnetics, elasticity). Both block-structured and unstructured grids are used to solve practical problems. The main difference between simulation of the thin film growth and the bulk crystal growth is that in the former case the computational domain can be considered fixed due to the small thickness of the epitaxial layer. Numerical study of the bulk crystal growth requires the use of either moving grids or a regeneration of the grid. The latter approach is attractive when one can exploit a quasi-stationary approximation of the growth processes (the characteristic time of the crystal shape changes is large compared to the hydrodynamic/thermal time).

A simulation of the crystal growth requires solution of the conjugated multidisciplinary problem. The key subproblem is the computation of the fluid flow coupled to the global heat transfer in the growth facility. Frequently, the global solution is used to specify boundary conditions for a smaller imbedded computational domain where a more elaborate physical model is considered.

**Conductive heat transfer** Two aspects of heat conduction in crystal growth facilities should be mentioned. Firstly, one need to account for the anisotropic thermal conductivity for certain crystals and for the solid blocks manufactured from pyrolytic graphite. In the first case the degree of anisotropy, being determined by the crystal composition and crystallographic symmetry, is usually not large. Pyrolytic graphite is obtained by pyrolysis of hydrocarbon gas at high temperature in the vacuum furnace and has a layered structure with highly ordered hexagonally arranged carbon atoms in planes and randomly oriented atoms in the perpendicular direction. The ratio of the values of the thermal conductivity in the different directions is 100-400, depending on the material quality. Whether anisotropy forces one to consider three-dimensional problem for geometrically two-dimensional configuration depends on the crystal symmetry and the orientation of its principal axes. For example, if the symmetry axis coincide with the [0001] axis of a hexagonal crystal such as SiC, the solution should be isotropic with respect to rotations around the axis and the two-dimensional formulation is valid.

Secondly, some parts of the facility could be porous medium. Powder source is used in a number of tecniques such as the metal flux method for growth of bulk GaN crystals from Li-Ga-N liquid phase [12], the ammonothermal method for GaN [13], sublimation growth of single crystals of wide bandgap semiconductors (SiC, AlN) [14,15]. Granular or fibrous medium is often used for insulation. The usual approach in the computation of the global heat transfer in the facility is to treat porous media using the effective thermal conductivity. For a given porous structure this quantity is a function of the pressure and the temperature that determine the relative contribution of the solid frame conduction, conduction through the filling medum (gas) and radiation to the total heat transfer. Experimental data being rather scarce, especially at high temperatures, the main problem is to formulate a model that could adequately extrapolate the effective thermal conductivity beyond the measured range of the pressures and temperatures [16,17]. The effective thermal conductivity could be two orders of magnitude smaller than that of bulk material that is evidently favorable for the use of porous media as insulation, but has a detrimental effect on the optimal heating of SiC powder source in the sublimation method [16]. The composition, porosity and the thermal conductivity of SiC powder vary during the growth process [18].

**Convective heat transfer**
*Melt/solution flow.* The great variety of the melt flow patterns observed in the crystal growth systems results from the highly different scales of crystals and dimensions of the growth equipment, variation of the melt properties (for example, the Prandtl number of silicon is three orders of magnitude smaller that that of some oxides) and numerous driving forces [19].

The flow in the facility for Bridgman growth method (horizontal, or directional solidification, and vertical as well as its modification VGF (Vertical Gradient Freeze)) is probably the simplest one to simulate. The flow is driven by natural thermal and concentration (double diffusive) convection. The melt flow is more complex in the case of the vertical Bridgman growth with centrifugation [20,21], being determined by the interaction of buoyancy with the Coriolis force. Rotational Bridgman method [22] adds forced convection and the free melt/gas interface (and thus the Marangoni effect). The forced flow in Czochraski (Cz) growth is caused by the crucible and the crystal rotation. An important phenomena to account for is a dynamic gas/melt interaction (the shear stress due to the gas flow can greatly exceed the Marangoni stress [23]). In Liquid Encapsulated Czochraski (LEC) method a presence of a third fluid - encapsulant - does not essentially increase the difficulty of the flow computations (the encapsulant flow is usually laminar), but could greatly complicate the mass transfer problem due to the numerous physical-chemical processes occurring in the three media and at the interfaces with poorly known rate constants and transport coefficients. In all the methods mentioned so far the melt/gas interface can be considered flat (except the relatively small meniscus region adjacent to the crystal) while in the floating zone (FZ) method it is usually highly curved [24].

Natural convection in the melt growth under microgravity could be induced by Marangoni effect as well as by non-stationary perturbations (g-jitter) caused by the crew motions and mechanical vibration of the satellite. The amplitude of the accelerations could be an order of magnitude larger than the gravity level at the orbit; sometimes the effect of the Coriolis force due to the satellite rotation around its axis and around the Earth is to be taken into account (see, for example, [25,26] and references therein). Resulting melt flow is usually has a detrimental effect on the crystal quality, in particularity, on the dopant distribution uniformity. However, vibrations of the prescribed character (progressive, rotational or swinging), amplitude and time variation could be used to control the growth process both at the Earth and at the orbital station [27].

One can judge whether the melt flow in the growth system should be laminar or turbulent by the Grashof or Rayleigh number and (rotational) Reynolds number. For example, natural convection is turbulent in the hydrothermal quartz growth: the Rayleigh number based on the autoclave *diameter* could as high as $4 \cdot 10^{11}$ [6].

In the case of mixed convection a winner is surely the silicon Czochralski growth due to the large size of the crystal and the crucible and the low melt viscosity. The turbulent melt flow in the crucible is non-steady and essentially three-dimensional [28]. An approach based on the Reynolds-averaged Navier-Stokes equations (RANS) could not predict with a sufficient for the engineering purposes accuracy such growth characteristics as the melt/crystal phase boundary shape or the oxygen concentration in the silicon crystal that are critical for the crystal quality [29,30].

On the other hand, at the present time Direct Numerical Simulation (DNS) of the melt flow is feasible for highly viscous melts such as oxides or in the simple model geometrical configurations [31]. DNS in the large-scale industrial equipment is hardly possible in the near future. For example, for Czochralski growth of 300 mm Si crystals the typical Reynolds numbers based on the crystal and crucible rotation rates are about $Re_{crys} \langle 10^5$ and $Re_{cruc} \langle 3 \cdot 10^5$, respectively. These values correspond to the Taylor-scale Reynolds numbers $Re_\lambda$ about 400 – 600 [32]. The highest value of $Re_\lambda$ = 1200 reported so far has been achieved for homogeneous turbulence in a periodic box using a spectral method with $4096^3$ ($\langle 69 \cdot 10^9$) grid points; for the indicated $Re_\lambda$ interval the size of the grid for this simple problem should contain about $1024^3$ ($\langle 1 \cdot 10^9$) – $2048^3$ ($\langle 8.6 \cdot 10^9$) cells [33,34]. Simulation of the oxygen transport in Czochralski silicon growth will require the use of even finer computational grids due to the high value of the Schmidt number [35,36].

Evidently, one should consider with caution the results of the so called "coarse DNS" or "quasi-DNS" computations. These approaches do not provide the necessary spatial resolution, but, as their advocates claim, reproduce the flow structures observed experimentally. Indeed, the coherent structures in the developed turbulent flow are known to weakly depend on the Reynolds number. However, in the crystal growth problems one is interested first of all in the flow behaviour near the crystal, the crucible and the free surface. This situation resembles the numerous attempts made about twenty years ago to simulate the separated viscous flows using Euler equations: such computations could reproduce the overall flow structure if a separation point/line is defined by the geometrical singularity, but could not give the skin friction and the heat flux.

Thus, at present, the most adequate methods for the modelling of the turbulent flow in industrial growth systems are Large Eddy Simulation (LES) and a hybrid method mixing the best features of LES and RANS approaches [37].

Since frequently the growth facility has an axisymmetric configuration, it is desirable to restrict 3D computations to the melt domain. Almost everybody fails to resist this temptation. This leads, however, to the additional artificial constraints on the flow – axisymmetric thermal boundary conditions, be they taken from the experiment or from the 2D global heat transfer computations. Moreover, these conditions usually are the stationary ones. More adequate approach is to extend 3D domain by adding some solid blocks: for example, in the simulation of Czochralski growth it is reasonable to include at least the crystal, quartz and massive graphite crucibles, probably the gas domain as well [38,39]. Computations based on such 3D model of the "growth zone" have shown, in particular, that temperature distribution at the melt/crucible boundary is essentially non-stationary and greatly deviates from the axisymmetric one [38].

*Gas flow.* Gas flow in crystal growth is laminar in most cases, except in the high pressure LEC of $A_3B_5$ semiconductors and is governed by the low-Mach number (hyposonic) Navier-Stokes equations [40]. These equations follow from the full compressible Navier-Stokes equations under the following assumptions [41]:
- Mach number of the flow is small;
- the hydrostatic compressibility is small (i.e. "shallow" convection only is considered);
- the characteristic time is large compared to the acoustic time.

Low–Mach number equations similar to incompressible flow equations do not describe acoustic phenomena: the only compressibility effect retained is the local thermal gas expansion. To obtain a well known Boussinesq approximation, one should additionally require the smallness of the relative temperature (density) variation.

In case of non-dilute mixtures a CFD problem is coupled to the mass transfer one [42,43]. Generally both homogeneous and heterogeneous chemical reactions are to be taken into account, the latter resulting in the highly nonlinear boundary conditions. When concentrations of active reagents in a carrier gas are high, homogeneous nucleation can occur. An advanced clustering model based on the moment-type equations was developed and implemented into the conventional heat and mass transport of multi-component reacting mixtures in ref. [44].

**Radiative Heat Transfer** Due to the high temperatures radiation is, as a rule, an important heat transfer mechanism and frequently a dominant one. It can be even used as a heating method as in floating zone growth with a mono- [45] or double-ellipsoid mirror furnace [46].

In the simplest case when all surfaces reflect diffusively and all solids and fluids present in the growth facility are either opaque or transparent, computation of grey radiation using configuration (view) factors (a so-called "surface-to-surface" model) [47] is all one needs.

Considerably more complex problem arises when some surfaces can exhibit specular reflection and/or the crystal or the melt should be treated as a semitransparent medium. Frequently one should also account for the spectral dependence of the optical properties and different values of the refraction index.

Semitransparency of the crystal significantly influences the structure of heat transport in the growth facility. The most evident manifestation of its effect is the variation of the of the shape of the phase boundary melt/crystal [48,49]. The intensity and the distribution of the radiative heat flux from the interface can cardinally change with the crystal length during the growth in the crystals with a high refraction index, the effect of the total inner reflection being responsible for this variation [50]. Other effects are also observed. For example, incorporation of microbubbles at the phase boundary results in the occurrence of diffusively transmitting and scattering boundary that could lead to the non-monotonic temperature variation in optical crystals [51]. The internal radiation in the melt usually is less important due to the higher absoption. However, it has been shown recently that neglecting the internal radiative heat transfer in the oxide melt $LiNbO_3$ results in the formation of the fictious spoke pattern [52]. Semitransparent properties of the $B_2O_3$ encapsulant in LEC have been accounting for in refs. [47,53].

Detailed comparison with experimental data [54] shows that the diffusion approximation of the radiative heat transfer via an additional "radiative" thermal conductivity used, for example, in ref. [55] to study $CaF_2$ melt growth, does not describe correctly temperature distribution in the system. Thus advanced models, such as the discreet exchange factor method [56] or the characteristics method [57] are needed.

A universal general approach (the ray tracing method) based on considering the generalized view factors for both the surface emitting and semitransparent volume elements in 3D arbitrary configurations was proposed in ref. [58]. The method allows an easy treatment of all the relevant optical phenomena in the system (absorption, scattering, refraction, diffuse and specular reflection) and is free from the "ray effect" unavoidable in all methods based on a discrete set of the radiation directions. However, it is quite time and memory consuming. Moreover, the ray emission model described the cited paper fails to account for semitransparent media with a large absorption coefficient, which, in turn, does not allow one to apply the method for multi-band computations. A new Ray Emission Model free of this drawback as well as the method extension to the multi-band radiative heat transfer have been suggested in [59]. The modified ray tracing method has a common deficiency of Monte Carlo methods: an unfavorable trade-off between the computation time and the predictions accuracy. A number of approaches to optimized the computational procedure including universal ones such as pencil ray tracing techniques, search acceleration routines etc. as well as specialized techniques including adaptive ones that are applicable to the restricted class of problems have been considered in [60]. An efficient modification of this method to axisymmetric problems have been reported in [59].

The effects of the specular reflection in the crystal growth problem have drawn lesser attention so far. It is usually assumed that the surface of the semitransparent crystal is covered by a vanishingly thin opaque film [49,56]. Moreover, sometimes to justify this simplification it is claimed to be valid for real systems due to parasitic deposition of impurities (it is funny, that such statements do not prevent the authors from indicating the refraction incorporation as a future work). Indeed, wall deposits are usual in the chemical vapour deposition reactors (and often semitransparent, increasing considerably the complexity of computations of radiative heat transfer, especially if the film thickness is comparable with the radiation wavelegth), but this phenomenon is hardly a universal one in the melt crystal growth processes. An opaque crystal surface means that the refraction index of the crystal has no effect on the heat transfer.

First computations accounting for the specular reflection from the crystal surface have been done for BGO crystal growth in the low temperature gradient Czochralski system using the characteristics method [57]. The shape of the melt/crystal boundary has been fixed in the cited paper; self-consistent computations have been reported in ref. [50]. It has been shown that incorporation of the specular reflection from the crystal inner surface, first of all from its conical part, (and thus taking into account the large value of the refraction index for bismute germanate crystal) allows one to reproduce the cardinal change of the interface shape with the length of the crystal observed in the experiments.

Another example of the importance of the specular reflection is the study of the global heat transfer in the industrial epitaxial Centura reactor using the modified ray tracing method [59]. Heating of the graphite susceptor and substrate in this reactor is provided by radiation from halogen lamps with golden reflectors. Three-band radiation model has been used to account for the spectral dependence of quartz optical properties. It has been found, in particularity, that quartz walls of the reactor chamber are heated not by the direct lamp radiation, but by the secondary radiation from the substrate having much lower temperature. Accuracy of the discrete ordinate model with diffusive reflections only turns out to be unsufficient in this problem.

**Phase Boundaries** The simulation of crystal growth from the melt includes a self-consistent determination of at least one unknown boundary. In Czochralski methods the interface melt/gas (melt/encapsulant, encapsulant/gas in LEC) can usually be assumed flat except a small meniscus region. To find this phase boundary in floating zone method one has to solve a coupled thermal-electromagnetic-hydrodynamic problem [24].

The boundary fluid/crystal is common for all growth methods. One can either track it explicitly or use one of the 'uniform' methods such as the enthalpy model [61-63], the level set approach [64,65], the phase-field model [66]. In the latter, for example, solid phase is considerd as a fluid with a very large viscosity [67]. This approach could be optimal for the growth of crystals with complex boundaries, such as dendritic solidification [68]. In simulation of the industrial growth of single crystal, the treatment of the melt/crystal boundary as a sharp phase interface is preferable.

For solidification of a "pure" substance (i.e. when mass diffusion effects can be neglected), the process is driven by the temperature gradient alone and is described by a classical Stefan problem. In the quasi-stationary formulation the interface should be fitted in such a way that the growth rate projection on the crystal pulling direction is constant.

The interface description is more difficult when alloy segregation or facetting of the crystal surface occur. Alloy segregation is a critical issue in the growth of bulk ternary III-V crystals [69]. These materials could provide lattice-matched substrates for epitaxial growth that eliminate the need for a graded buffer layer and thus reduce the production costs and increase the life-time of semiconductor devices due to the lower density of the misfit dislocations. The difficulty of growth of such ternary compounds as GaInAs and GaInSb stems from the large solidus - liquidus separation of the phase diagram and thus a small value of the In segregation coefficient. The rejected solute accumulates at the interface and is transported by diffusion and convection, the latter having a major effect on the quality and the composition uniformity of the grown crystals [69,70]. Various defects such as striations or cellular structure in ternary In-containing crystals leading to polycrystallinity when the solute concentration exceeds a critical value [71] are often observed. The temperature dependence of the solute diffusivity can have a significant effect on the onset of instability [72]. A predictive description of the phase boundary should account for a number of local coupled phenomena (heat and mass transfer, melt flow, interfacial kinetics, constitutional undercooling, morphological instability of the crystal/melt interface).

In the case of facetting of the crystal surface, the melt/crystal boundary does not follow the melting point isotherm, but coincides with a crystallographic plane. An interface undercooling could be large (for example, almost 7 K for silicon [73] and up to 20 K in bismuth germanate growth [74]). It is usually claimed that the faceting is harmful for crystal quality [75] and thus should be avoided by keeping thermal gradient above a critical value. However, recently large nearly perfect crystals of bismuth germanate have been grown with the fully facetted solidification front [76]. A numerical treatment of the partially facetted interface has been discussed recently in ref. [77].

**Electromagnetics** The degree of coupling of the electromagnetic problem to other phenomena depends on the growth method in question. Magnetic field used to damp the turbulence fluctuations in Czochralski method or to provide a controlled action on the melt flow should be computed self-consistently with the flow or considered given, depending on the value of the magnetic Reynolds number. Radio frequency (RF) heating is only weakly coupled to the thermal problem via temperature-dependent material properties in the sublimation, Bridgmen and some other crystal growth methods. On the other hand, in the floating zone method it is coupled to the flow that determines the free surface shape and even to the dopant segregation at the growth interface via local electric resistivity [24].

**Assessment of Crystal Quality** The growth process simulations themselves could provide the growth rate and the composition distribution only. The ultimate aim being the crystal quality, one has also to analyse the thermal stress, formation of point defects and dislocations in the crystal and their evolution during the growth and the post-growth processing.

When the crystal deformation is pure elastic, the stress can be computed for any growth stage independently. The thermoelasicity problem is a three-dimensional one even for an axisymmetric crystal, except the case of a special orientation of the principal crystal axes [78-80]. Stress caused by the temperature gradients (as well as by the compositional inhomogeneity in ternary compounds induced by segregation) could result in the formation of cracks in brittle crystals [81]. Compositional strain in ternary crystals containing Ga and In could greatly exceed thermoelasic one since the tetrahedral radii of the two substituting atoms Ga and In differ by 12 per cent [70].

When the stress level exceeds a critical value, a plastic creep occurs in the ductile crystals. Models of the dislocation evolution usually exploit the plastic strain rate dependence on the deviatoric stress and dislocations parameters ( the density, the velocity, the Burgers vector, etc.) [82] and an equation for the evolution of the dislocation density [75,83,84].

Dislocations are universal in the sense of being present in the practically all kinds of crystals. Other defects are specific for some crystals only: for example, the formation of twins and low angle boundaries is really important in binary $A_3B_5$ compounds such as GaAs and InP while polytype inclusions are observed in SiC, incorporation of microbubbles [85] is deleterious in the growth of oxides, halides and other optical crystals. The incorporation of the intrinsic point defects into a growing crystal and their evolution, including the formation of voids and oxygen precipitates, are of paramount importance for the silicon crystals. These processes essentially depend on the melt/crystal interface shape, the ratio of the growth rate to the axial temperature gradient and the presence of impurities, i.e. on the peculiarities of the heat and mass transfer. Simulation of defects in Si crystals has been reviewed recently in ref. [86].

The degree of coupling of the crystal quality assessment procedure to the simulation of the growth process itself varies greatly. For example, the compositional uniformity and the concentration of bubbles in the optical crystals could be determined simultaneously with the growth rate. Computations of the thermal and compositional stress in the case when no plastic deformation occurs, being stationary problems due to the high sound velocity in the solids, could be performed as post-processing. Evolution of dislocations or voids and oxygen precipitates are nonsteady problems that, however, could be considered independently. An example of the most closely coupled problem is probably impurity segregation in the floating zone method with inductive heating due to the effect of the impurity distribution on the melt and crystal electric conductivity.

Evidently, one has to search for a compromise between the model completeness and tractability. Rephrasing the well-known quotation, the development of a model is finished not when there are no more relevant phenomena to incorporate, but when one can not exclude an effect without compromising the model.

## SOFTWARE

**Requirements** The focus of numerical simulation of crystal growth is moving now from universities and academia to industry. There are several reasons for that: a great number of available commercial and public-domain generic CHT and CFD codes [87]; cheap high power hardware; industry's reluctance to reveal the proprietary information to the outside consulting partners; in-house operation allows one to use simulation routinely in the everyday work. Still, "it is certainly not the case that commercial products can be used in general engineering design without support from fluid mechanics specialists" [88], to say nothing of numerous "non-CFD" complications. CHT and CFD are certainly ones of the most mature computational technologies evolved from the academic research into the widespread industrial application. However, it is recognized that a non-expert is rarely able to apply them successfully to the industrial problems. CFD is considered as an uncertain discipline and a knowledge-based activity. To cure the situation, EU has launched a huge project QNET-CFD [88] (over 40 organizations from 11 states) with the aim not to perform actual research, but to assemble and arrange *existing* knowledge encapsulating CFD use in the different industrial sectors (external aerodynamics, combustion, chemical and civil engineering, environment, turbomachinery flows) and to establish the best practice guidelines. Numerical simulation of crystal growth, being based on CHT and CFD, inherits all their concerns and adds a number of own problems such as unknown phase boundary, facetting of crystal surface, anisotropic crystal properties, thermal stress, formation and evolution of point and extended defects during the growth as well as during the post-process cooling. Industry certainly needs customized multidisciplinary simulators that hide from the user intricacies of the numerical issues and allow the engineer to concentrate on the problem to be solved.

There are two approaches to the development of such growth simulators. One can either "wrap" general purpose code(s) or design a "dedicated" simulator. It is intersting to note that, as far as the authors know, only codes for epitaxial growth can be found in the first group (PHOENICS-CVD [89], CVD-Module [90]) while all the bulk growth simulators (FEMAG [91], Cape simulators [92], CrysVUN [93], STHAMUS [93], Virtual Reactor [94], CGSim [90]) as well as some CVD simulators (Cape products [92], PROCOM [95]) are developed from a scratch. Somewhat exaggerated differences in the requirements for software used in academia and industry are given in Table 1 [96].

Table 1.
Comparison of numerical simulation in academia and industry

| Feature | Academia | Industry |
|---|---|---|
| Aim | Insight | Optimization |
| Geometry | Simple | Complex |
| Domain | Single or a few blocks | Complete multi-block system |
| Physics | One or a few phenomena | Multidisciplinary |
| Properties | Constant or analytical | Real |
| Platform | Workstation up to mainframe | Workstation or cluster of WS |
| Execution time | Not critical | Up to hours |
| Run-time tuning | Possible | Unacceptable |
| Robustness | Not critical | Mandatory |
| User's experience | Up to high | None up to average |
| Foolproof input | Optional | Mandatory |
| On-line help | Optional | Mandatory |
| Documentation | Optional | Mandatory |

**Verification and validation** The necessary stages of the code development are verification (an assessment of the correctness of the model implementation) and validation (an assessment of the adequacy of the model to the real world) [97]. Sometimes it is stated that nature of the code (research, pilot, production) is determined by its maturity with respect to the validation level [98]. This seems to be oversimplification. If one needs a single criteria (ideally, quantifiable) to estimate the practical usefulness of simulation, the best choice is probably *the reliability of a computer prediction* [99]. It should be stressed, however, that this parameter characterizes not the *code* itself, but the *simulation*, being depending on the adequacy of the model and the accuracy of the computations as well as on the particular aim of the simulations. Evidently, the same results could be considered successful if one is interested in unveiling some trend - and unsatisfactory if the goal is to find, for example, the optimal size of some specific element of the growth facility.

*Verification.* Verification is primarily a mathematical issue [97]. The five major sources of errors in the numerical solution have been listed in ref. [100]: insufficient spatial discretization convergence; insufficient temporal discretization convergence; insufficient convergence of an iterative procedure; computer round-off; computer programming errors. The errors of the last type are the most difficult to detect and fix when the code executes without an obvious crash yielding "moderately incorrect results" [100]. The study reported in ref. [101] revealed a surprisingly large number of such faults in the tested scientific codes (in total over a hundred both commercial and research codes regularly used by their intended users).

Verification is performed by comparison of the numerical results with "highly accurate" (benchmark) solutions. There are three main sources of such solutions: 1) exact analytical solutions; 2) benchmark solutions of ordinary differential equations; 3) benchmark solutions of partial differential equations.

The usual use of ODE solution is based on exploiting the symmetry properties: one can solve an essentially one-dimensional problem (for example, having spherical symmetry) using a general three dimensional grid. There are a few well known multidimensional benchmark solutions such as laminar convection in a square cavity [102,103], laminar flow over a backward-facing step with heat transfer [104], axisymmetric heat and mass transfer in chemical vapour deposition (CVD) of silicon in a rotating disk/stagnation flow reactors [43].

Additional tests could be constructing via a so called "method of manufactured solutions" by inserting an arbitrary analytic function into the equations and evaluating for what source terms and boundary conditions this function will be an exact solution. This approach is attractive since it could be applied to complex multidimensional domains while allowing direct assessment of the solution accuracy. However, it usefulness depends on the similarity of the guessed "exact" solution to the typical solution of the real problem in question. An availability of the analytic exact solution greatly reduces the CPU time needed for testing of nonstationary codes: one can use an exact solution for the specification of initial and, if necessary, (nonsteady) boundary conditions and avoid the need of using large time integration intervals.

*Validation.* Validation is the second step in the assessment of the software quality. It should be stressed that verification should not be skipped: successful validation alone does prove the reliability of the code due to limitness of a set of validation cases, a "graphical" comparison with experimental data in the most cases and a possible cancellation of multiple errors that can give can an impression of a success.
The validation of the heat transfer and crystal growth computations in the complete facility could be extremely difficult due uncertainty in properties of different materials, the necessity to quantify a complete set of test conditions and provide a comprehensive set of measurements of major parameters with the error estimates. Thus a decomposition of the system into an hierarchy of subproblems could be useful [100]. A few model problems inspired by crystal growth are listed in ref. [105]. Among them are an "annular configuration" (a cylindrical container with a small coaxial cylindrical heater) that can be used to study thermocapillary convection driven by the radial temperature gradient and a "half zone" [106] that emulate the liquid bridge of the melt in the floating zone method.

The major efforts in crystal growth are directed to the characterization of the obtained crystals. There are numerous experimental approaches to assessment of the crystal quality in regards to surface and bulk defects [107]. Various optical microscopy and interferometry techniques such as phase contrast microscopy (PSM), differential interference contrast microscopy (DISM), atomic force microscopy (AFM), and scanning tunneling microscopy (STM) can visualize and measure growth steps of nanometer height. In some cases such as protein crystallization it is even possible to monitor, due to the large size of macromiolecules and large characteristic times, the behaviour of an individual "growth unit" on the crystal surface [108-110]. Generation and spatial distribution of dislocations, other lattice defects, chemical inhomogeneities in crystals can be analyzed using X-ray topography, laser-beam tomography, cathodoluminescence tomography, micro-focused X-ray fluorescence and so on.

There are, however, serious technical difficulties in the systematic *in situ* measurements of the temperature field and flow visualization during the crystal growth. One is the reasons is the extremal parameters frequently used in crystal growth (for example, growth of GaN single crystals from solutions of nitrogen in liquid gallium is carried out at the temperature about 1800 K under $N_2$ pressure up to 15 000 bar [111,112]).

Semiconductor melts are opaque, still, flow visualization is possible using at least two approaches. The first one is the introduction of radioactive tracers [113], the second one, called real-time X-ray radioscopy (radiography) [114,115] is based on the dependency of the X-ray absorption on the material density. Thus for the melt of pure substance this method gives temperature distribution. The case of the alloy melts is more complex since both the temperature and the concentration variations contribute to the variation of the density, and, hence, absorption. When the absorption coefficient of one component of the binary solution is much greater that that of the other, one gets an approximate visualization of the concentration distribution [116]. X-ray radioscopy could be used for *in situ* monitoring of the melt-crystal phase boundary [117]. Recently a confocal scanning laser microscope with an infrared image furnace has been used to observe the evolution of the morphology of silicon crystal with increasing the growth rate and to measure the melt undercooling at the facet [73].

A unique method for direct measurement of the melt-crystal interface temperature based on the transparency of the bismuth germinate $Bi_4Ge_3O_{12}$ crystal and opaqueness of the corresponding melt allowed the authors of ref. [74] to study *in situ* the undercooling of the phase boundary in the Axial Heat Processing (AHP) method. It should be noted that this method is good for model validation due to the simple geometry of the growth facility, practically suppressed convection (the high melt viscosity, the small melt height) and ability to measure directly the temperatures of the melt and crystal boundaries and thus exclude the necessity of the computation of the global heat transfer in the facility [118].

In addition to the flow visualization using above described and traditional techniques (interferrometry, holography, addition of solid tracers), only few reliable *in situ* temperature measurements have been reported. The measurements of the fluid temperature are relatively easy for the low temperature growth. Thus, authors of ref. [119] have determined the axial temperature gradients in the phosphoric acid solution in the conditions used for the growth of $AlPO_4$ and $GaPO_4$ crystals at approximately 500 K. Free surface temperature fluctuations caused by the thermocapillary convection in liquid metal (Si, Mo) bridge have been measured in a number of studies both at the Earth and reduced gravity (see, for example, [120-122]).

Probably the most advanced *in situ* high temperature measurements have been reported in refs. [28,123]. The authors were able to measure the temperature and temperature fluctuations within the silicon melt during industrial crystal growth using thermocouples and optical sensors. The superiority of the latter has been established, since signals from thermocouples do not adequately reproduce the high frequency part of the turbulence spectrum. The authors also have developed sensors for *in situ* measurement of the oxygen concentration both in the melt and in the gas phase.

The difficulties of the study of real growth processes *in situ* force one to exploit *physical* simulation of crystal growth. Thus, in ref. [124] silicon oil has been used to study the flow in the Czochralski crucible. An outstanding research has been reported recently in ref. [125]. To simulate the processes in the large scale silicon growth, the authors exploited an eutectic mixture InGaSn with the melting point about 285 K. Tuning the crucible size, the magnetic induction and the "crystal" and crucible rotation rates, the authors were able to achieve the simultaneous equality of four criteria (the Grashof number, the Hartmann number and two Reynolds numbers, determined by the "crystal" and the crucible rotation) in the physical experiment to that in the industrial 200 mm silicon growth. Still, there a few distinctions in simulation from the real process: to approximately model the melt heat losses by the radiation from the free surface, special coolers have been desighned and installed; it is hardly possible to emulate Marangoni effect and the shear stress, caused by the gas flow over the melt free surface; the "crystal" part of the InGaSn melt upper boundary was flat while in silicon growth it is highly curved.

Fortunately for CHT, the shape of the fluid/crystal phase boundary is quite sensitive to the local heat transfer and this information, being available in majority of the growth experiments (either via visualization of the growth striations or using intentional "pulsed" doping for interface evolution monitoring), is certainly one of the major "measuring sticks" for code validation.

**Could Software be User-Friendly?** It has been claimed that terms such as "user-friendly" or "easy-to-learn" are ambiguous because they are subjective and thus unverifiable [126]. On the other hand, they can be measured in the relative units - one can easily compare two codes using the time needed to master the code operation by an uninitiated user or the time required for the specification of the geometry and the problem parameters. To make the code attractive to the industry user, developers should

1. Use robust algorithms that does not require run-time monitoring and tuning
2. Minimize the user actions required for the problem specification
3. Use units, variables and control parameters specific for the growth method in question]

Ideally, the code should be a black-box one that requires no intervention by the user. The price of the robustness is efficiency. To find a compromise, different forms of adaptivity should be exploited. Grid adaptation to the solution can easily be automated, using as a stopping criteria (in steady problems) either the specification of the finest grid size [127] or, which is more properly (but more tediously), an error estimation [128]. For the most time consuming part - iterative solution of large sparse systems of equations - one can use an adaptive polyalgorithm (an ordered set of iterative methods from the fast, but the least robust to the most robust slow one) with an automatic method switching [129].

Examples of the item 2 are automatic block detection in the geometry entered by the user and automatic updating geometry and grid regeneration during the growth process (caused by the crystal shape evolution, the heater, the inductor or the boat movement).

Complete automation could be extremely difficult. Probably the most time (and money [130]) consuming part of the problem statement is a geometry specification. Unfortunately, geometry import of a CAD model usually requires a number of user actions generically referred to as *CAD data repair* [131]. For example, many wide-spread formats for CAD models do not provide the neibourhood relations for model entities. Thus, the topology of the model must be reconstructed and this procedure is non-trivial due to the erroneous gaps between the neigbour elements [132]. Often primary reason for numerous geometry and topology inconsistencies in CAD data (such as, for example, an edge whose path intersects itself or a face with edges that do not form a closed loop) is a difference between tolerances used by a CAD designer and tolerances needed for grid generation [130]. Thus at present a manual repair of imported CAD data (probably assisted by interactive repair and defeaturing tools [133]) seems to be inevitable for complex geometry.

To conform to the item 3, the developers' goodwill is only needed.

## OPTIMIZATION OF THE GROWTH PROCESS AND EQUIPMENT

The aims of optimization usually are to increase the crystal size and the uniformity of crystal properties, decrease the number of defects and production costs. A straightforward scaling of the growth equipment according the crystal dimension does not work due to the nonlinearity of the underlying physical phenomena. Note, for example, that the Reynolds number scales linear with the characteristic length, the Grashof number as a third power while the Marangoni number does not depend directly on the linear scale.

Optimization of the operating conditions of bulk crystals and thin films growth is now in its infancy while that of growth equipment is still in the prenatal state. A few known examples of the process optimization use a small number of the control parameters (such as the heater(s) power/position in melt growth [75,134] or mass flow rates and susceptor rotation rate in CVD [135]) and thus probably do not suffer from the ill-posedness of the inverse problems to be solved. When the number of control parameters is large, one is forced to use a regularization of some kind [9] as, for example, in the control of the composition variation in MOVPE growth of ternary $Al_xGa_{1-x}As$ films [136] or in the optimization of the crucible design for SiC bulk crystal growth [14,137]. Probably the most advanced example of the crystal growth optimization is an application of the adjoint method to the solution of the inverse problem for the optimal boundary heat flux distribution in the directional solidification and Bridgman method [138] and optimal crystal surface temperature distribution in Czochralski growth [139].

High fidelity direct problem solvers being not fast enough, one is often forced to use, at least at the early optimization stages, some "surrogate" models [140]. Such low fidelity models could be either physically motivated (a reduced spatial dimension of the problem) or derived as black-box models via multivariable approximation (regression methods, neural networks, kriging etc. [141]). Note that in the latter case the numerical and experimental data could easily be combined in the optimization process.

The parametric geometric modeling, being essentially morphing of a few curves/surfaces, severely restricts the search space. To increase the power of the optimization, topological changes in the system configuration should be allowed. To summarize, there is still a long way to the development of software for optimal design of the entire crystal growth system. Obviously, experience in multidisciplinary optimization should be borrowed from more (computationally) mature industry sectors such as aerospace engineering [141,142].

## AN EXAMPLE: BULK SIC CRYSTAL GROWTH

Silicon carbide is a perspective material for high-temperature, high-power, high-frequency electronic and optoelectronic devices [14,15]. It is superior to conventional semiconductor materials such as Si, GaAs, InP, having larger operating temperature range, higher critical breakdown field, high resistance to radiation. SiC has numerous applications both as material for solid state devices themselves (high power rectifiers and switches, transistors for microwave technology etc.) and as substrate for optoelectronic devices. For example, SiC substrates greatly outperform sapphire ones for high brightness nitride-based light emitting diodes in all respects except cost. Thus growth of high quality large silicon carbide crystals is of paramount importance.

**Sublimation growth** In the sublimation method a single crystal is grown from the vapour phase in a closed crucible, the transport being provided by a suitable temperature gradient between the powder charge and the seed. The heart of the growth facility is the crucible with SiC powder at the bottom and a seed at the top (Figure 1). The temperature gradient is provided by the radio-frequency inductive (sometimes resistive) heating. Unlike chemical vapor deposition, the sublimation technique does not allow a separate control of silicon and carbon supply, which makes the precise control of the temperature distribution in the growth chamber necessary.

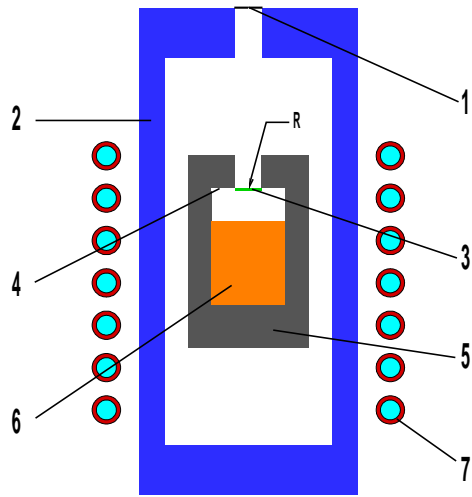

Figure 1. Schematic view of a sublimation growth system. Basic elements: 1) pyrometric window, 2) thermal insulation, 3) substrate, 4) the upper part of the growth system, 5) the crucible, 6) powder charge, 7) inductor coils.

The thermal field simultaneously affects many various factors, such as growth rate at the seed, SiC powder sublimation rate, etching of growth crucible graphite walls and SiC deposition on the walls, evolution of the SiC crystal shape during the growth, graphite and silicon inclusions into the crystal, thermoelastic stress and the dislocation generation, etc. The requirements for control of these factors are known to come frequently into contradiction with each other. For instance, high growth rate requires a considerable axial temperature gradient in the crucible, whereas this favors graphitization of the growth surface and stimulates dislocation and micropipe generation due to enhanced thermoelastic stresses. As a result, a reasonable compromise should be found to meet all the requirements.

**Mathematical model**  To simulate the whole growth process, one should compute the temperature and concentration distribution in the system and monitor the evolution of the crystal front, possible parasitic deposits at the crucible wall and powder source. The process parameters usually are the pressure, the heating power and the inductor position.

To formulate a comprehensive model, one has to consider all major physical-chemical phenomena relevant to this method such as resistive or RF heating ; conductive, convective and radiative heat transfer; mass transfer in gas and porous medium; heterogeneous chemical reaction at catalytic walls and on the surface of powder granules; deposit formation at the crucible walls; formation of elastic strain and dislocations in the growing crystal; evolution of the crystal and deposit shape, including partial facetting of the crystal surface; evolution of the composition ("graphitization") and porosity of the powder. The crystal growth rate being rather small (0.3-1.0 mm/h), the problem can be solved using a quasi-stationary formulation.

The Joule heat source distribution is determined by solution of the Maxwell equations the linear isotropic nonuniform medium in the frequency domain. Global heat transfer analysis includes conductive, convective and radiative transfer modeled using the configuration factors. To treat nonideal solid-solid contacts in the real growth equipment, thermal gaps should be allowed that can be described using the effective heat transfer coefficient that accounts for radiation, gas conduction and partial solid direct-contact conduction. The Darcy law is used to relate the velocity and the pressure distribution in the powder.

The vapour phase in sublimation SiC growth consists of Si, Si$_2$C, and SiC$_2$, diluted by an inert gas, usually argon. Conjugate mass transfer in the gas and the powder using Hertz-Knudsen fluxes in the formulation of the boundary conditions for the species concentrations gives the growth rate at the catalitic surfaces. The kind of growing crystal is determined by the type of the surface and local concentration of gas phase components.

**Software** The code "Virtual Reactor" (VR) [14,137,143] for crystal growth by sublimation has been developed as a tool for industry engineers. VR has an easy-to-learn interface that allows the user either to describe the geometry manually or to import a CAD file. All geometric and process parameters entered by the user are checked automatically to belong to the corresponding interval of admissible values. The code has a number of features aimed at minimizing the user's efforts such as

- automatic initial identification of the blocks from the wireframe geometry
- automatic identification of the types of the inner boundaries
- one-time specification of the whole growth process
- automatic updating of the boundaries of the crystal and deposit blocks
- (almost) automatic processing of the topological changes of the computational domain (formation of new blocks and boundaries)
- automatic (re)generation of the unstructured grid in the the modified and new blocks

Specification of the growth process includes formulation of the time variation of the pressure, the RF power and the inductor position. Unstructured grid is generated block-wise using Delaune algorithm, an advancing front method or their combination. Non-matched grids in the neighbour blocks are allowed. At the each step a number of subproblems is solved subsequently. Computation of the thermal stress and dislocation density is implemented as a post-processing procedure. The code has a vast extendable data base of material properties and the built-in visualization module.

Transfer to the next step includes the propagation of the crystal and deposit boundaries, identification of new blocks and boundaries (if needed), the movement of the inductor (if specified by the user) and unstructured grid generation in the new and altered blocks. A special optimization procedure for the growing front advancement has been developed that eliminate the effect of the numerical noise in the growth rate distribution and allows a stable evolution of the crystal and deposit shape and a monitoring of the topological changes in the computational domain. As an example, the crystal and deposit shapes along with the isotherms are shown in Figure 2 after 5 and 40 hours from the start of the growth process. The case when the seed is placed on the seedholder is considered.

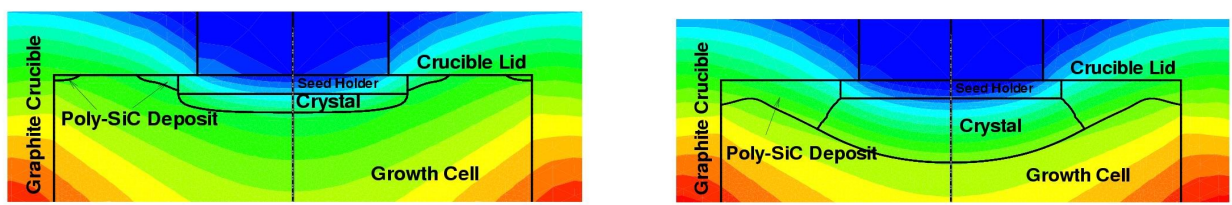

Figure 2. Evolution of the shape of growing SiC crystal and poly-SiC deposit after 5 (left) and 40 (right) hours.

VR can be used for the optimization of both crystal growth process and equipment. The latter is illustrated by Figure 3 where the crucible geometry and isotherms are presented. Initial temperature distribution is shown in Figure 3 (a) while results of the crucible shape optimization aimed at the minimization of the axial temperature gradients and both axial and radial gradients in the growth zone are displayed in Figure 3 (b) and (c), respectively.

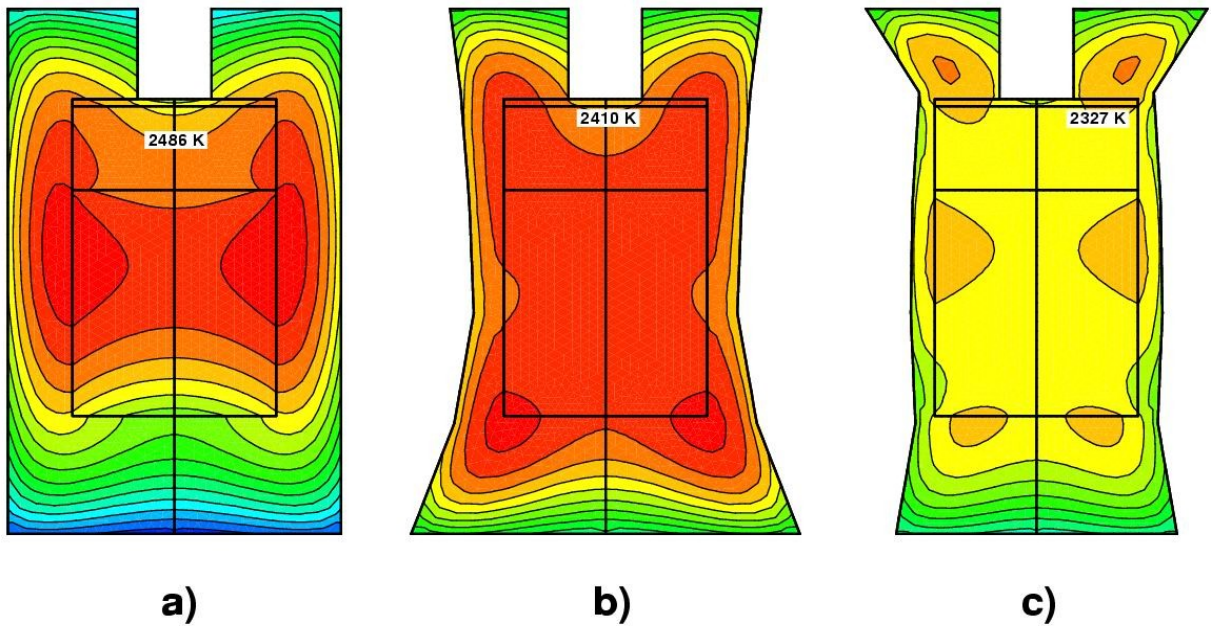

Figure 3. Optimization of the crucible design: initial (a); minimization of the axial temperature gradients (b); minimization of axial and radial temperature gradients (c).

## CONCLUSIONS

An attempt is made to review the heat transfer and related problems encountered in the simulation of single crystal growth. Peculiarities of conductive, convective and radiative heat transfer in the different melt, solution, and vapour growth methods are discussed. The importance of the adequate description of the optical crystal properties (semitransparency, specular reflecting surfaces) is stresses. Treatment of the unknown phase boundary fluid/crystal as well as problems related to the assessment of the quality of the grown crystals and their coupling to the heat transfer/fluid flow problems is considered. Differences between the crystal growth simulation codes intended for the research and for the industrial applications are indicated. The problems of the code verification and validation are discussed; a brief review of the experimental techniques for the study of heat transfer and flow structure in crystal growth is presented. The state of the art of the optimization of the growth facilities and technological processes is discussed. An example of computations of the heat transfer and crystal growth of bulk SiC crystals by sublimation method is presented.

## ACKNOWLEDGEMENTS

The author is grateful to M.V. Bogdanov, D.Kh. Ofengeim, S.K. Kochuguev, A.V. Kulik, M.S. Ramm, A.M. Serkov, A.V. Tsirulnikov, V.S. Yuferev, and I.A. Zhmakin for collaboration and useful discussions.